\documentclass[aps,pre,preprint]{revtex4-1}
\usepackage{graphicx}
\usepackage{amsmath}
\usepackage{dcolumn}
\usepackage{bm}
\begin{document}
\title{On equivalence of high temperature series expansion and coupling parameter series expansion in thermodynamic perturbation theory of fluids} 
\author{A. Sai Venkata Ramana} %
\affiliation{Theoretical Physics Division, Bhabha Atomic Research Centre, Mumbai, 400 085, India}
\email[asaivenk@barc.gov.in]\\  
\date{\today}
\begin{abstract}
The coupling parameter series expansion and the high temperature series expansion in the thermodynamic 
perturbation theory of fluids are shown to be equivalent if the interaction potential is pairwise additive.
 As a consequence, for the class of fluids with the potential having a hardcore repulsion, if the hard-sphere fluid
 is chosen as reference system, the terms of coupling parameter series 
expansion for radial distribution function, direct correlation function and Helmholtz free energy follow a scaling 
law with temperature. The scaling law is confirmed by application to square-well fluids.
\end{abstract}
\maketitle 

\section{Introduction}
Perturbation series for Helmholtz free energy of a fluid at a given temperature and density 
was first written down by Zwanzig\cite{zwanzig}. The basic idea is to
 split the total potential energy($V$) of the fluid containing N-particles into a reference part($V_{ref}$) and a perturbation part ($V_{per}$).
 Assuming that $V_{per}$ is small, the term $exp(-\beta V_{per})$ in the canonical partition function of the fluid is expanded in a series
 of $-\beta V_{per}$. Where $\beta = 1/k_bT$; $k_b$ being Boltzmann constant and T being temperature.
This results in a cumulant expansion for Helmholtz free energy. 
The series is also known as high temperature series expansion(HTSE).
The perturbation series of Zwanzig is general in the
 sense that it doesn\rq{}t restrict to pairwise additive interactions between the fluid particles. However, if the 
interactions are pairwise additive, Zwanzig showed that the first order term in the series depends on radial 
distribution function(RDF) of  the reference system and 
the second order term depends on correlation functions up to fourth order. An approximation to the second order term 
was obtained by Barker and Henderson\cite{barker1}. It was not possible to calculate beyond second order as the higher 
order correlation functions are unknown.
 
The perturbation series for the Helmholtz free energy can be derived in alternative ways\cite{barker2, hansen} using 
coupling parameter or strength parameter ($\lambda$) which defines the strength of perturbation.
 The total potential energy of an imaginary fluid is written as, $V_{\lambda} = V_{ref} + \lambda V_{per}$.
$\lambda = 0$ gives the reference fluid potential energy and $\lambda=1$ gives potential energy of the fluid under 
consideration. One way of arriving at the perturbation series is to differentiate the Helmholtz free energy 
of the imaginary system w.r.t. $\lambda$. The derivative depends on ensemble average of $V_{per}$ w.r.t. the imaginary 
fluid i.e., $<V_{per}>_{\lambda}$. $<V_{per}>_{\lambda}$ is expanded as a Taylor series around $\lambda=0$ and then
the equation is integrated within the limits $\lambda=0$ and $\lambda=1$. This is called as $\lambda$-expansion and
 the terms are exactly the same as the HTSE. 

Another way of obtaining the series, which works only in the case of a pairwise additive potential between the particles, is as follows:
The Helmholtz free energy of the system is expanded as a Taylor series in $\lambda$ around $\lambda = 0$. 
 In the case of a pairwise additive potential, it turns 
out that the Taylor series of Helmholtz free energy depends on the Taylor series of the pair correlation function with 
respect to $\lambda$ at $\lambda = 0$. This expansion of Helmholtz free energy is referred as coupling parameter series 
expansion(CPSE).
 In order to calculate $n^{th}$ order term in this expansion, $(n-1)^{th}$ derivative of pair correlation function is 
supposed to be known. The first order term in the CPSE can be directly seen to be equal to first order perturbation 
term in HTSE. 
The method to calculate higher order terms (including the second order term in an accurate way) was not known 
for a long time. In 2006, Zhou\cite{zhou1} developed a numerical method to calculate the
 higher order derivatives of RDF using the integral equation theory (IET). Zhou\rq{}s method was, however, 
computationally intensive and the calculated derivatives have numerical fluctuations and require further smoothing 
procedures. Recently, we developed a much simpler method which doesn\rq{}t have those problems\cite{sai1,sai2}.
 The ability to calculate higher order terms, which was not possible with the earlier HTSE enormously 
improved the accuracy of predictions of the theory.

In a few recent papers\cite{zhou2, zhou3,zhou4, zhou5,zhou6}, it has been argued that the HTSE and CPSE are actually different 
beyond the first order term. It also has been stated that, only in the special case of hard sphere reference system and only 
for reduced temperature equal to 1, the two series are equal. However, the conclusions were not based on any
 rigorous results and the arguments were heuristic. Apart from this, a function having two different series expansions contradicts the 
 uniqueness of power series expansion.
 This raises questions about the validity of the conclusions made in the papers\cite{zhou2, zhou3,zhou4, zhou5,zhou6}.
 In view of this, we re-examine the equivalence of CPSE and HTSE and prove rigorously that both the expansions are equivalent.
 Even though this result is not very surprising, the 
equivalence of CPSE and HTSE leads to interesting scaling relations in the case of fluids with interaction potential having
 a hardcore repulsion(for 
example, the square well (SW) fluid) if the hardcore repulsive part is taken as reference. 
 For such fluids, the HTSE becomes a power series in $\beta$. Hence , the terms of CPSE also scale with powers of $\beta
$. As a consequence, once the terms of CPSE are calculated along a particular isotherm for such fluids, the structure 
and thermodynamic properties of those fluids at other temperatures can be obtained using the scaling relations.

In section II, we prove the equivalence of CPSE and HTSE and in section III we describe the consequences of the equivalence of HTSE and CPSE. The paper is concluded in section IV.

\section{The HTSE and the CPSE}
Consider a fluid of N interacting particles in a volume $\Omega$ at reduced temperature T. Reduced units ($\epsilon/k_B 
= \sigma =1$, where $\epsilon$ is the well depth and $\sigma$ is hard sphere diameter) are used throughout the paper. 
 The Zwanzig\rq{}s expansion or HTSE for Helmholtz free energy $F$ of the system is given by
\begin{equation}
F = F_{ref} + \sum_{n=1}^{\infty} \frac{\omega_n }{n!}(-\beta)^{n-1} \label{z1}
\end{equation}
where 
\begin{eqnarray}
\omega_1 &=& \langle V_{per} \rangle_0 \\               \label{z2}
\omega_2 &=& \langle V_{per}^2 \rangle_0 - \langle V_{per} \rangle^2_0 \\ \label{z3} 
\omega_3 &=& \langle V_{per}^3 \rangle_0 -3\langle V_{per}^2 \rangle_0 \langle V_{per} \rangle_0 +2\langle V_{per} \rangle^3_0 \\ \label{z4}
& &........... \nonumber \\
\omega_s &=& s!\sum_{\{n_j\}} (-1)^{\Sigma n_j -1 }\left(\Sigma n_j -1\right)! \prod_{j=1}^\infty\frac{1}{n_j!}\left( \frac{ \langle V_{per} \rangle_0^j}{j!}\right)^{n_j} \label{z5}
\end{eqnarray}
The subscript $0$ implies that the averages are with respect to the reference system ensemble.
In the Eq.(\ref{z5}), the ${\{n_j\}}$ implies that the summation is a multiple sum over all 
possible $n_j$\rq{}s such that the 
constraint $\sum_0^\infty j n_j = s$ is satisfied. Here $n_j$\rq{}s are positive integers. It can be seen that for all 
$j>s$, $n_j$'s are zero. Above series for $F$ doesn\rq{}t require a pairwise additive interaction potential. However, 
in the case of a spherically symmetric pairwise additive potential $u(r)$, the $\omega_1$ simplifies to

\begin{equation}
\omega_1 = \frac{2\pi N^2}{\Omega} \int_0^\infty u_{per}(r)g_0(r)r^2 dr \label{z6}
\end{equation}
where $u_{per}(r)$ is the perturbation part of $u(r)$ when separated as $u(r) = u_{ref}(r) + u_{per}(r)$. $g_0(r)$ is 
the radial distribution function of the reference system. $r$ is the radial distance from the origin. $\omega_2$ has 
been shown to depend on correlation functions up to fourth order. An approximation to $\omega_2$ has been derived by Barker and 
Hendersen\cite{barker1}. 

In the $\lambda$-expansion\cite{hansen} and the CPSE, as mentioned in the introduction, an imaginary fluid is considered 
with total potential energy given by $V_{\lambda} = V_{ref} + \lambda V_{per}$. Differentiating $F(\lambda)$ $w.r.t.$ $
\lambda$ and integrating between limits $\lambda =0$ to $\lambda=1$ gives

\begin{equation}
F = F_{ref} + \int_0^1 \langle V_{per} \rangle_{\lambda}d\lambda \label{c1}  
\end{equation}
The subscript $\lambda$ in the above equation implies that the average is $w.r.t.$ the ensemble with potential energy 
$V_{\lambda}$. Expanding $\langle V_{per} \rangle_{\lambda} $ around $\lambda = 0$ in a Taylor series and substituting $
\lambda = 1$, we get the HTSE. In the case of pairwise additive potentials, above Eq.(\ref{c1}) can be re-written in 
terms of pair correlation function ( $\rho^{(2)}_{\lambda}(1,2)$) of the imaginary system i.e.,
 \begin{equation}
 F = F_{ref} + \frac{1}{2}\int_0^1 \!\!\! d\lambda \int_\Omega\!\!\! d1d2 u_{per}(1,2)\rho^{(2)}_{\lambda}(1,2)  \label{c2}  
 \end{equation}
Following Hansen\cite{hansen}, we denote the position vector ${\vec r_i}$ of $i^{th}$ particle by $i$ to simplify
notation. $d1$ denotes the volume element $d{\vec r_1}$ around ${\vec r_1}$. 
 Expanding  $\rho^{(2)}_{\lambda}(1,2)$ in a Taylor series around $\lambda = 0$ and substituting in above 
Eq.(\ref{c2}), we get the CPSE for $F$ given by
  
 \begin{equation}
 F = F_{ref} + \frac{1}{2}\int_0^1 \!\!\! d\lambda \int_\Omega\!\!\! d1d2 u_{per}(1,2)\left(\rho^{(2)}_{0}(1,2) +  \lambda \left. \frac{d\rho^{(2)}}{d\lambda}(1,2)\right|_0 +  \left. \frac{\lambda^2}{2!}\frac{d^2\rho^{(2)}}{d \lambda^2}(1,2)\right|_0 + ........... \right) \label{c3}  
 \end{equation}
   If the system is homogeneous and isotropic, $\rho^{(2)}_{\lambda}(1,2)$ becomes equal to $\rho^2 g_{\lambda}(|1-2|)$ 
   and the integral in Eq.(\ref{c3}) simplifies. Here $\rho$ is the density of the system and $g_{\lambda}(r)$ is the RDF of the imaginary system.

\subsection{Equivalence of HTSE and CPSE}
It can be easily seen that first order terms in both the series are same; as mentioned above, in a homogeneous system,
$\rho^{(2)}_{\lambda}(1,2)$ becomes equal to $\rho^2 g_{\lambda}(|1-2|)$ and 
\begin{equation}
\frac{1}{2}\int_0^1 \!\!\!d\lambda \int_\Omega\!\!\!d1d2 u_{per}(1,2)\rho^{(2)}_{0}(1,2) = \omega_1\\ 
\end{equation}

In order to see the equivalence of second term of CPSE with that of HTSE, consider the definition
 of $\rho^{(2)}_{\lambda}(1,2)$ i.e.,

\begin{equation}
\rho^{(2)}_{\lambda}(1,2) = N(N-1)\frac{1}{Z_N(\lambda)}\int_\Omega d{\vec r^{(N-2)}}exp(-\beta V_{\lambda}) \label{e1}
\end{equation}
where
\begin{equation}
Z_N(\lambda) = \int_\Omega d{\vec r^{(N)}}exp(-\beta V_{\lambda}) \label{e2}
\end{equation}

Here $\int_\Omega d{\vec r^{(N)}}$ implies $\int_\Omega...\int_\Omega d1d2...dN$

Now, 
\begin{eqnarray}
\frac{d\rho^{(2)}_{\lambda}}{d \lambda} &=& \frac{N(N-1)}{Z_N(\lambda)}\int_\Omega d{\vec r^{(N-2)}}(-\beta V_{per})exp(-\beta V_{\lambda}) \nonumber \\
& &- \frac{N(N-1)}{Z_N^2(\lambda)}\int_\Omega d{\vec r^{(N-2)}}exp(-\beta V_{\lambda})\int_\Omega d{\vec r^{N}}(-\beta V_{per})exp(-\beta V_{\lambda}) \label{e3}
\end{eqnarray}

Substituting the above equation in second order term of CPSE we get,
\begin{eqnarray}
\frac{1}{2}\int_0^1 \!\!\! d\lambda \int_\Omega\!\!\! d1d2 u_{per}(1,2) \lambda \left. \frac{d\rho^{(2)}}{d\lambda}(1,2)\right|_0 &=& \frac{N(N-1)}{2Z_N(0)}\int_\Omega d{\vec r^{N}}(-\beta V_{per})exp(-\beta V_{ref})u_{per}(1,2) \nonumber \\
&-& \frac{N(N-1)}{2Z_N^2(0)}\int_\Omega d{\vec r^{N}}u_{per}(1,2)exp(-\beta V_{ref})\nonumber \\
& &\times \int_\Omega d{\vec r^{N}}(-\beta V_{per})exp(-\beta V_{ref}) \nonumber \\
     \label{e4}
\end{eqnarray}
Consider the first term in the RHS of the above equation. It can be seen that the integral remains invariant if $(1,2)$ 
is replaced by any other possible pair and there are exactly $N(N-1)/2$ possible pairs. The same argument holds for the second term on the RHS of above Eg.(\ref{e4}). Thus the above equation becomes,
 
\begin{eqnarray}
\frac{1}{2}\int_0^1 \!\!\! d\lambda \int_\Omega\!\!\! d1d2 u_{per}(1,2) \lambda \left. \frac{d\rho^{(2)}}{d\lambda}(1,2)\right|_0 &=& \frac{1}{Z_N(0)}\int_\Omega d{\vec r^{N}}(-\beta V_{per})exp(-\beta V_{ref})V_{per} \nonumber \\
\mbox{ }\mbox{ }& &- \frac{1}{Z_N^2(0)}\int_\Omega d{\vec r^{N}}V_{per}exp(-\beta V_{ref}) \nonumber \\
& &\times\int_\Omega d{\vec r^{N}}(-\beta V_{per})exp(-\beta V_{ref}) \nonumber \\
& =& -\beta \langle V_{per}^2\rangle_0 + \beta \langle V_{per}\rangle_0^2\nonumber \\
& =&  -\beta \omega_2/2 \label{e5}
\end{eqnarray}

Now to prove the equivalence of  a general term of CPSE to that of HTSE, we first define the following:
 \begin{eqnarray}
P_0 &=& P = \int_{\Omega} d3..dN exp(-\beta V_{\lambda}) \nonumber \\ 
P_1 &=& \frac{dP}{d\lambda} = \int_{\Omega} d3..dN (-\beta V_{per})exp(-\beta V_{\lambda}) \nonumber \\ 
P_2 &=&  \frac{d^2P}{d\lambda^2} = \int_{\Omega} d3..dN (-\beta V_{per})^2 exp(-\beta V_{\lambda}) \nonumber \\ 
  & &.......................................\nonumber \\
P_s &=& \frac{d^sP}{d\lambda^s} = \int_{\Omega} d3..dN (-\beta V_{per})^s exp(-\beta V_{\lambda}) \label{e6}
 \end{eqnarray}
and
 \begin{eqnarray}
Z_0 &=& Z_N(\lambda) = \int_{\Omega} d1..dN exp(-\beta V_{\lambda}) \nonumber \\ 
Z_1 &=& \frac{dZ_N(\lambda)}{d\lambda} = \int_{\Omega} d1..dN (-\beta V_{per})exp(-\beta V_{\lambda}) \nonumber \\ 
Z_2 &=&  \frac{d^2Z_N(\lambda)}{d\lambda^2} = \int_{\Omega} d1..dN (-\beta V_{per})^2 exp(-\beta V_{\lambda}) \nonumber \\ 
  & &.......................................\nonumber \\
Z_s &=& \frac{d^sZ_N(\lambda)}{d\lambda^s} = \int_{\Omega} d1..dN (-\beta V_{per})^s exp(-\beta V_{\lambda}) \label{e7}
 \end{eqnarray}
Here $d1$ implies $d {\vec r_1}$ and so on.
It can be easily seen that 

\begin{eqnarray}
\frac{1}{N(N-1)}\rho^{(2)}_{\lambda}(1,2) &=& \frac{P_0}{Z_0} \nonumber \\
\frac{1}{N(N-1)}\frac{d \rho^{(2)}_{\lambda}(1,2)}{d \lambda} &=& \frac{P_1}{Z_0} - \frac{P_0}{Z_0^2}Z_1 \nonumber \\
\frac{1}{N(N-1)}\frac{d^2 \rho^{(2)}_{\lambda}(1,2)}{d \lambda^2} &=& \frac{P_2}{Z_0} - 2\frac{P_1}{Z_0^2}Z_1 
                               +2\frac{P_0}{Z_0^3}Z_1^2 - \frac{P_0}{Z_0^2}Z_1^2  \label{e8}
\end{eqnarray}

To get the $s^{th}$ ($s \ge 1$) derivative  of $\rho^{(2)}_{\lambda}(r_1,r_2)$, the $s^{th}$ derivative of $\frac{P_0}{Z_0}$ is
required which in turn requires the the $s^{th}$ derivative of $\frac{1}{Z_0}$. Since $Z_0$ is a well behaved function,
 it can be expected that $s^{th}$ derivative of $\frac{1}{Z_0}$ w.r.t. $\lambda$ exits.
The $s^{th}$ derivative of $\frac{1}{Z_0}$ is \cite{allen}

\begin{eqnarray}
\frac{d^s}{d\lambda^s}\left(\frac{1}{Z_0}\right) &=& s!\sum_{\{n_j\}}(-1)^{\alpha}\frac{\alpha!}{Z_0^{\alpha + 1}} \prod_{j=1}^\infty\frac{1}{n_j!}\left( \frac{ Z_j}{j!}\right)^{n_j} \nonumber \\ 
\alpha &=&  \sum_{j=1}^\infty n_j \label{e9}
\end{eqnarray}
where $\sum_{\{n_j\}}$ is again a multiple sum over $n_j$\rq{}s such that the constraint $\sum_{j=1}^\infty jn_j = s$ is 
satisfied.
Using Leibnitz formula for calculating the $s^{th}$ derivative of a product of two functions we get
\begin{equation}
\frac{1}{N(N-1)}\frac{d^s\rho^{(2)}_{\lambda}(1,2)}{d\lambda^s} = \sum_{r=0}^s \frac{s!}{(s-r)!r!} P_{s-r}\frac{d^r}{d\lambda^r}\left(\frac{1}{Z_0}\right) \label{e10}
\end{equation}
The $(s+1)^{th}$ term in CPSE(Eq.(\ref{c3})) is,
\begin{equation}
\Gamma_{s+1} = \frac{1}{(s+1)!}\frac{1}{2}\int_{\Omega}d1d2 u_{per}(1,2)\left. \frac{d^s\rho^{(2)}_{\lambda}(1,2)}{d\lambda^s}\right|_0 \label{e11}
\end{equation}
Substituting Eq.(\ref{e10}) in Eq.(\ref{e11}) and after some algebra, the $(s+1)^{th}$ term in CPSE becomes
\begin{eqnarray}
\Gamma_{s+1} &=& \frac{1}{(s+1)!}\sum_{r=0}^s \frac{s!}{(s-r)!} \frac{\langle (-\beta V_{per})^{s-r+1} \rangle_0}{-\beta}\sum_{\{n_j\}}(-1)^{\alpha}\alpha!\prod_{j=1}^\infty\frac{1}{n_j!}\left[\frac{\langle (-\beta V_{per})^{j} \rangle_0}{j!}\right]^{n_j} \\ \label{e12}
&=& \frac{1}{(s+1)!}\frac{-1}{\beta}\sum_{r=0}^s s!\sum_{\{n_j\}}(-1)^{\alpha}\alpha! \nonumber \\
& &\times\prod_{\substack{j=1 \\ j\neq s-r+1}}^\infty\frac{1}{n_j!}\left[\frac{\langle (-\beta V_{per})^{j} \rangle_0}{j!}\right]^{n_j}\frac{(n_{s-r+1}+1)(s-r+1)}{(n_{s-r+1}+1)!}\left[\frac{\langle (-\beta V_{per})^{s-r+1} \rangle_0}{(s-r+1)!}\right]^{n_{s-r+1}+1} \label{e13}
\end{eqnarray}
Now consider a set of $\{n_j'\}$s such that, for all $j\neq s-r+1$, $n_j' = n_j$ and for $j= s-r+1$, $n_j' = n_j + 1$.
Then, 
\begin{equation}
\sum_{j=1}^\infty jn_j' = \sum_{j=1}^\infty jn_j + (s-r+1) = s+1 \label{e14}
\end{equation}
and 
\begin{equation}
\alpha'=\sum_{j=1}^\infty n_j' = \sum_{j=1}^\infty n_j + 1 = \alpha+1 \label{e15}
\end{equation}
Using Eq.(\ref{e14}) and Eq.(\ref{e15}) in Eq.(\ref{e13}), we get,
\begin{equation}
\Gamma_{s+1} = \frac{1}{(s+1)!}\frac{-1}{\beta}\sum_{r=0}^s (n_{s-r+1}')(s-r+1)s!\sum_{\{n_j'\}}(-1)^{(\alpha'-1)}(\alpha'-1)!\prod_{j=1}^\infty\frac{1}{n_j'!}\left[\frac{\langle (-\beta V_{per})^{j} \rangle_0}{j!}\right]^{n_j'} \label{e16}
\end{equation}
where $n_j'$s satisfy the constraint of Eq.(\ref{e14}).
It can be seen easily that $\sum_{r=0}^s (n_{s-r+1}')(s-r+1) = s+1$, as all $n_j$s for $j>s+1$ must be zero for the
 constraint of Eq.(\ref{e14}) to be satisfied. Thus Eq.(\ref{e16}) becomes
\begin{eqnarray}
\Gamma_{s+1} &=& (-\beta)^s\sum_{\{n_j'\}}(-1)^{(\alpha'-1)}(\alpha'-1)!\prod_{j=1}^\infty\frac{1}{n_j'!}\left[\frac{\langle (V_{per})^{j} \rangle_0}{j!}\right]^{n_j'} \\ \label{e17}
&=& (-\beta)^s\frac{\omega_{s+1}}{(s+1)!}
\end{eqnarray}
which is nothing but $(s+1)^{th}$ order term in HTSE.

\section{Fluids with Hardcore Repulsion}
Consider a homogeneous fluid with particles interacting via a pairwise additive spherically symmetric potential.
Let the potential $u(r)$ have a hardcore repulsion($u_{ref}(r)$) and a slowly varying attractive part($u_{per}(r)$) whose value is 
small compared to $k_bT$ so that it can be treated as a perturbation. i.e.,
\begin{eqnarray}
u(r) &=& \infty \mbox{  } r \le r_0 \nonumber \\
     &=& u_{per}(r) \mbox{  } r > r_0
\end{eqnarray}
If the hardcore part is taken as the potential of the reference system, the ensemble 
averages $w.r.t.$
the hard-sphere reference system become independent of temperature. As a result, the $\omega_i$s in Equations(\ref{z2}-
\ref{z5}) become independent of temperature. Hence the HTSE becomes a power series in $\beta = 1/{k_bT}$. This result is
well known\cite{hansen}. Since it has now been shown that the terms of CPSE are equivalent to those of HTSE,
 the terms of CPSE also scale with powers of $\beta$. $s^{th}$ term of the series is proportional to $\beta^{s-1}$
as can be seen from Eq.(\ref{e17}).

It can also be seen that the derivatives of 
 RDF and the direct correlation function(DCF) also scale with powers of $\beta$.
Consider $s^{th}$ derivative of $\rho_{\lambda}(1,2)$.
It can be seen from  Eq.(\ref{e10}) that each term in the summation in the R.H.S. depends upon the following term:
\begin{equation}
\frac{P_{s-r}}{Z_0^{\alpha+1}}\prod_{j=1}^{\infty}Z_j^{n_j}
\end{equation}
which can be re-written as follows
\begin{eqnarray}
\frac{P_{s-r}}{Z_0^{\alpha+1}}\prod_{j=1}^{\infty}Z_j^{n_j} &=&\frac{P_{s-r}}{Z_0}\prod_{j=1}^{\infty}\left[\frac{Z_j}{Z_0}\right]^{n_j} \nonumber \\
&=& \frac{P_{s-r}}{Z_0} \beta^r(\langle(-V_{per}) \rangle_0^{n_1} \langle (-V_{per})^2 \rangle_0^{n_2}\langle(-V_{per})^3 \rangle_0^{n_3}......) \label{f1}
\end{eqnarray}
When evaluated at $\lambda = 0$ with $u_{ref}(r)$ as that of hard-sphere, it can be seen from the Eq.(\ref{e6})
that $P_{s-r}$ depends on $\beta^{s-r}$. Since the averages in above Eq.(\ref{f1}) become independent of temperature,
the total dependence on $\beta$ of each term in the R.H.S. of Eq.(\ref{e10})  would be  $\beta^s$. Thus,
\begin{equation}
\left.\frac{d^s\rho^{(2)}_{\lambda}(1,2)}{d\lambda^s}\right|_{\lambda=0} =\rho^2\left.\frac{d^sg_{\lambda}(r)}{d\lambda^s}\right|_{\lambda=0}\propto \beta^s \label{f2}
\end{equation}
Also, since $\left.\frac{d^sg_{\lambda}(r)}{d\lambda^s}\right|_{\lambda=0}\propto \beta^s$, its fourier transfrom also
would be proportional to $\beta^{s}$ i.e.,

\begin{equation}
\left.\frac{d^s\hat{g_{\lambda}}(k)}{d\lambda^s}\right|_{\lambda=0}=\left.\frac{d^s\hat{h_{\lambda}}(k)}{d\lambda^s}\right|_{\lambda=0}\propto \beta^s
\end{equation}
where $h_{\lambda}(r)=g_{\lambda}(r)-1$ is the total correlation function and $\hat{h}_{\lambda}(k)$ is its fourier
transform. The Ornstein-Zernike equation for the imaginary fluid in fourier space is given by\cite{hansen},

\begin{equation}
h_{\lambda}(k) = \frac{c_{\lambda}(k)}{1-\rho c_{\lambda}(k)} \label{oze}
\end{equation}
Using similar arguments given in the case of $\left.\frac{d^sg_{\lambda}(r)}{d\lambda^s}\right|_{\lambda=0}$, it can 
be seen that 
\begin{equation}
\left.\frac{d^sc_{\lambda}(r)}{d\lambda^s}\right|_{\lambda=0}\propto \beta^s \label{f3}
\end{equation}

Thus from the above discussion, it can be seen that, for the class of fluids with hardcore repulsion, apart from
 the Helmholtz free energy series, the Taylor series for $c(r)$ and $g(r)$ obtained by expanding 
around $\lambda = 0$ also becomes power series in $\beta$.
 As a result, if the coefficients of the series are obtained along one isotherm,
 then the whole phase diagram can be obtained provided the series converges. To obtain the coefficients of the series 
which are related to the derivatives of $g_{\lambda}(r)$  and $c_{\lambda}(r)$ at 
$\lambda = 0$, few recently developed methods are available as discussed in the introduction{\cite{zhou1,sai1,sai2}}.

\begin{figure}
\includegraphics[scale=0.5,angle=-90]{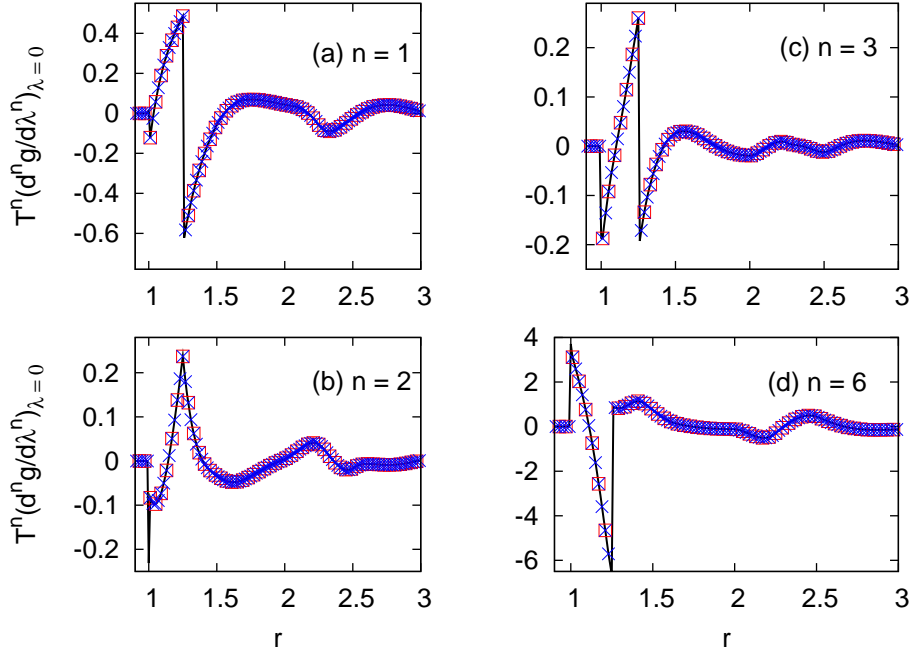}
\caption{\label{1} (Color Online) Scaled derivatives of g(r) of SW fluid of width $0.25$ at reduced density $\rho =0.8$ at
various reduced temperatures. Solid line: T = 0.9, hollow squares: T = 0.6, crosses: T =1.5}
\end{figure}

\begin{figure}
\includegraphics[scale=0.5,angle=-90]{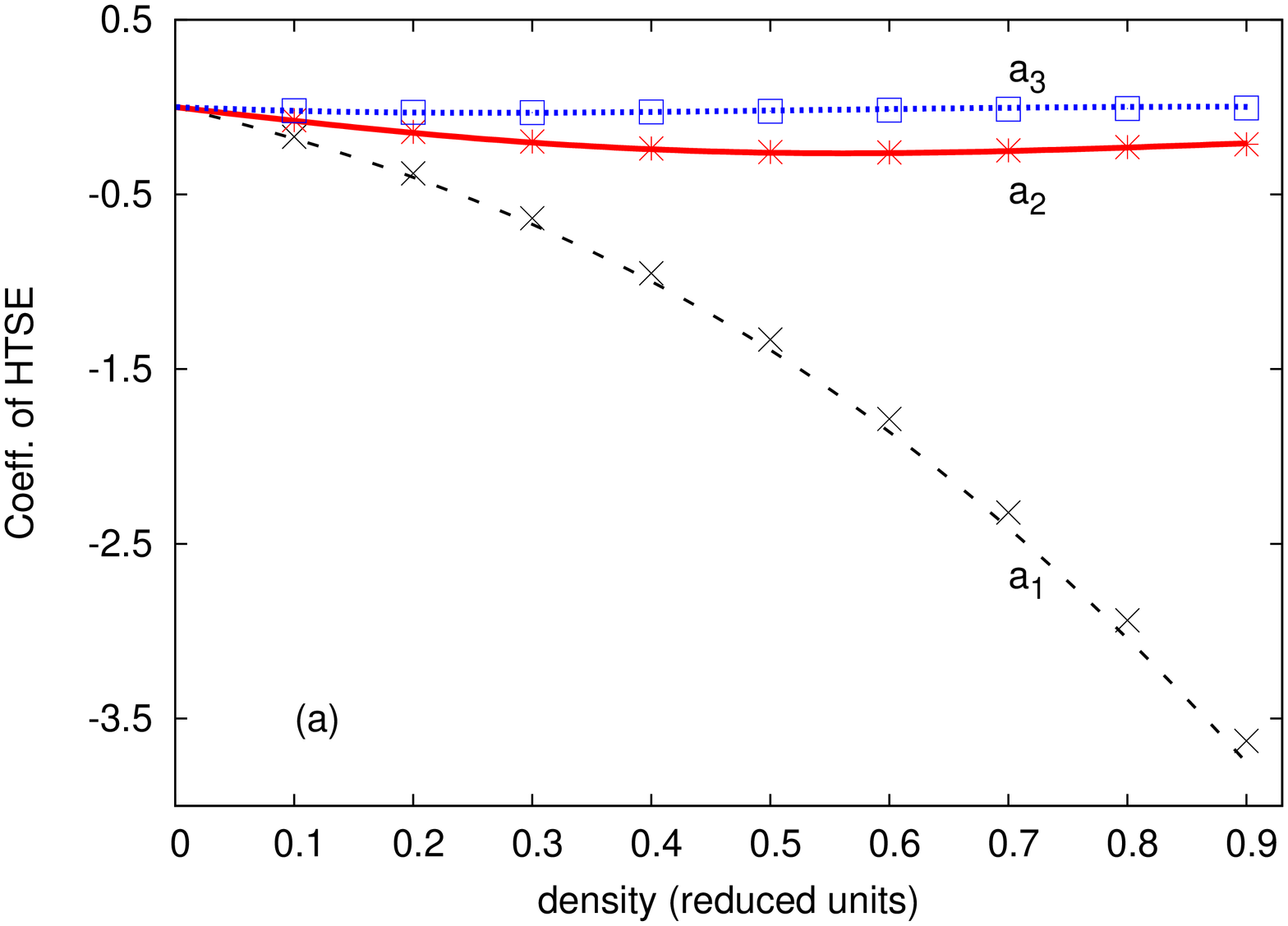}
\caption{\label{2} (Color Online) First three coefficients of HTSE for SW fluid of width $1.2$. Lines are present results, symbols are simulation
data\cite{zhou2}. $a_1, a_2$ and $a_3$ are first three coefficients of HTSE for Helmholtz free energy per particle respectively. }
\end{figure}

\begin{figure}
\includegraphics[scale=0.5,angle=-90]{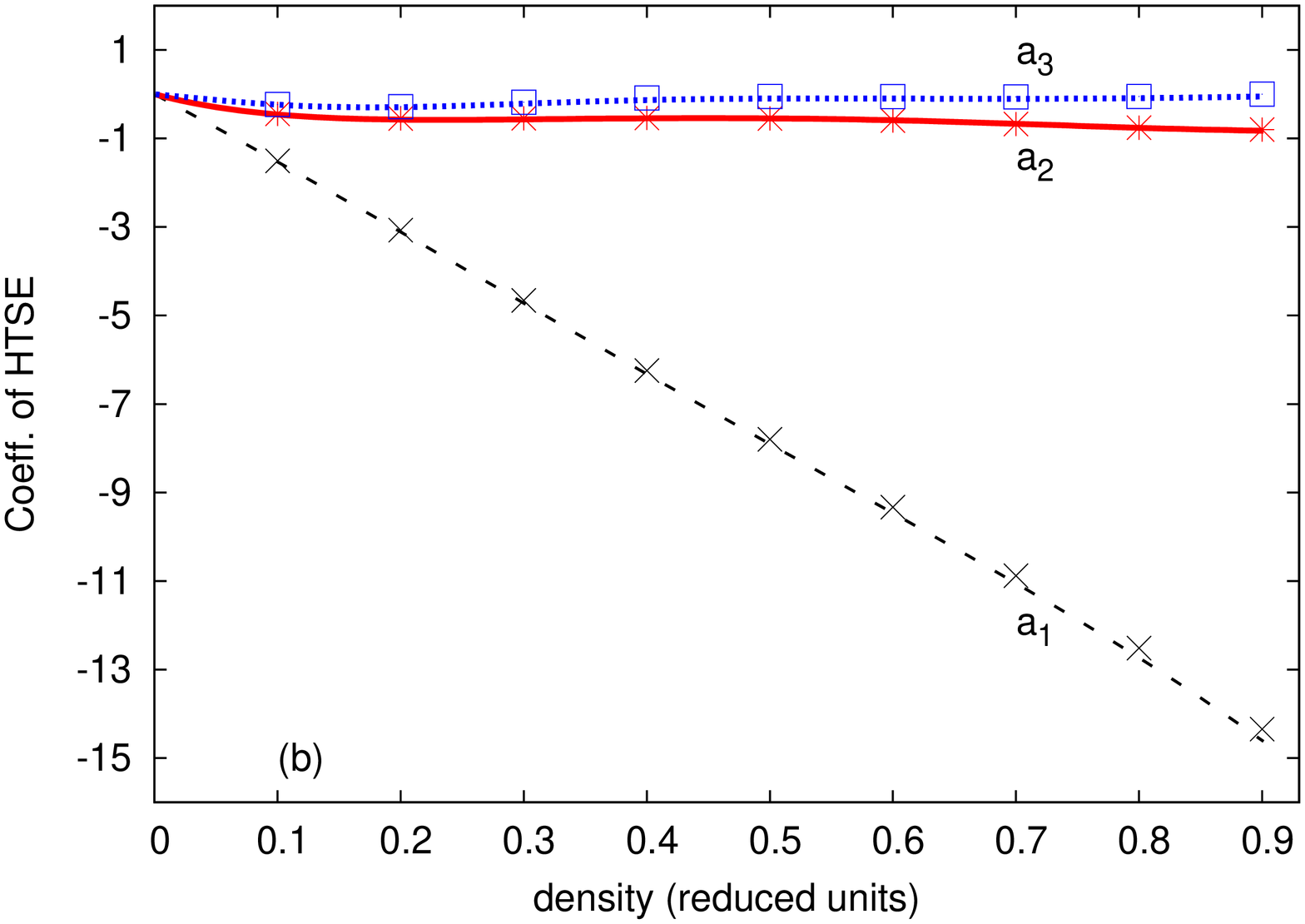}
\caption{\label{3} (Color Online) First three coefficients of HTSE for SW fluid of width $2.0$. Lines are present results, symbols are simulation data\cite{zhou2}.$a_1, a_2$ and $a_3$ are first three coefficients of HTSE for Helmholtz free energy per particle respectively.  }
\end{figure}

To test the scaling relations derived above, we applied the method discussed in Ref.\cite{sai2} to square well(SW) 
fluids. The calculation method used is exactly the same as in Ref.\cite{sai2} and hence is not repeated here. We obtained 
the derivatives of $g(r)$ of SW fluid of width $0.25$ at various reduced temperatures(T) and
 plotted $T^n\frac{d^ng_{\lambda}(r)}{d\lambda^n}|_0$ for $n=1,2,3$ and $6$ in Fig.(\ref{1}).
  It can be seen from Fig.(\ref{1}) that the
curves corresponding to each derivative at different temperatures are coinciding after scaling.
Also we obtained the first three coefficients of the HTSE for SW fluids of width $1.2$ and $2.0$ using the same 
method of Ref.\cite{sai2}. In the case of SW fluid of width $1.2$, the coefficients of HTSE are obtained by 
calculating the terms of CPSE at $T = 0.7$ and scaling with temperature according to $T^n\Gamma_{n+1}$ where
 $\Gamma_{n}$ is $n^{th}$ order term in the CPSE for Helmholtz free energy per particle. In the same way, 
the coefficitents of HTSE for SW fluid of width $2.0$ are obtained by calculating the terms of CPSE at $T=2.0$
 followed by scaling. The results are compared with simualtion data\cite{zhou2} in
Fig(\ref{2}) and Fig(\ref{3}). It can be seen that there is an
 excellent match between simulation data and present calculations which confirms the scaling relations.

\section{Conclusion}
We have shown rigorously that the CPSE and HTSE are equivalent when the interaction potential is pairwise additive.
 As a result, we have shown that for fluids with interaction potential having a hardcore
repulsion, if the hard-sphere fluid is taken as the reference system, the terms of CPSE and the derivatives of RDF and
DCF w.r.t. the coupling parameter follow a scaling relation with temperature. 
We also have confirmed the scaling relations through applications to square well fluids.
Thus, for such fluids, if the derivatives of RDF are
known along an isotherm, the structural and thermodynamic properties can be obtained at any other temperature provided
the series is convergent at that temperature.


\section{Acknowledgments}
I thank Dr. D. Gaitonde and Dr. D. Biswas for stimulating discussions.

\end{document}